\documentclass[conference]{IEEEtran}
\IEEEoverridecommandlockouts
\usepackage{cite}
\usepackage{amsmath,amssymb,amsfonts}
\usepackage{algorithmic}
\usepackage{graphicx}
\usepackage{textcomp}
\usepackage{xcolor}
\usepackage{url}
\usepackage{graphics}
\usepackage[skip=5pt]{caption}
\usepackage{bbm}
\usepackage{tikz}
\usetikzlibrary{arrows,calc,positioning}
\usetikzlibrary{positioning}
\usetikzlibrary{arrows.meta}
\usetikzlibrary{shapes,arrows}
\usetikzlibrary{shapes.geometric}
\def\BibTeX{{\rm B\kern-.05em{\sc i\kern-.025em b}\kern-.08em
    T\kern-.1667em\lower.7ex\hbox{E}\kern-.125emX}}
\begin{document}

\title{Activity Recognition Using mm-Wave Radar and Deep Learning: Prayer Tracker Case Study}

\author{\IEEEauthorblockN{Karim Saifullin, Sajid Ahmed, \IEEEmembership{Senior Member,~IEEE,} and Mohamed-Slim Alouini, \IEEEmembership{Fellow,~IEEE,}}
\IEEEauthorblockA{\textit{Electrical and Computer Engineering (ECE)}\\{Computer, Electrical and Mathematical Sciences and Engineering (CEMSE)} \\
\textit{King Abdullah University of Science and Technology (KAUST)},\\
Thuwal, 23955-6900, Kingdom of Saudi Arabia\\
karim.saifullin,sajid.ahmed,slim.alouini@kaust.edu.sa}}
\maketitle

\begin{abstract}
The issue of privacy has gained significant attention in recent times. Many real-world applications increasingly require the use of sensitive data, such as in surveillance or tracking and assistance systems. To address these concerns, we propose a framework based on mm-wave radar technology that not only meets privacy requirements but also provides the necessary capabilities for these systems, including reliable current position tracking, sequence tracking, and feedback to the user. While the use of radar technology for surveillance purposes is gaining momentum, there has been no research to date on its application for prayer tracking and assistance systems. Furthermore, there is a lack of comprehensive research that covers all aspects of implementing such a system. Proposed approach offers a versatile solution that can be applied to a broad range of scenarios. Instead of utilizing raw I-Q data, we addressed the challenge of classification based on point cloud information generated by the conventional processing chain of the frequency-modulated continuous wave radar. This information contains corresponding range, reflection amplitude, Doppler and angular values. We have developed and compared different machine-learning classification algorithms to identify the most effective one. Our findings reveal that the convolutional neural network ResNet achieves the best results, with accuracy rates reaching up to 95.4 percent when applied to unknown data. The demonstration video of the developed system can be viewed at the following link: \url{https://youtu.be/PnpGQZWqCr4}.
\end{abstract}

\section{Introduction}
Privacy awareness has become a crucial topic in recent times, as advancements in technology have made it easier to capture and store personal data. Some European countries, including Germany and France, have implemented strict privacy laws that restrict or forbid the use of video surveillance. In these countries, video cameras are generally only permitted in public places when there is a significant threat to security, and when it is necessary to prevent crime and safeguard public safety. These laws can present challenges for law enforcement agencies, but they are essential for protecting individuals' privacy rights.

In addition to law enforcement, there are many other applications where video monitoring is required while respecting people's privacy. For instance, in hospitals, video cameras may be used to monitor patients in their beds and bathrooms to help prevent falls and provide immediate assistance when needed. However, it is crucial to ensure that patients' privacy is protected and that any monitoring is carried out in compliance with relevant privacy regulations. Furthermore, some conservative societies may have reservations about the use of video cameras, particularly in worship places, where gender separation is observed. In such cases, special consideration must be given to the privacy and modesty of worshippers, particularly in areas designated for women. 

Overall, while video surveillance has many benefits, it is essential to balance the need for surveillance with the need to protect individuals' privacy rights. Radar can be used for range, angle, and velocity measurement; these capabilities of radar can be used for activity monitoring without recording any video. In situations where camera-based technology is not permitted due to privacy concerns, radar technology can serve as an alternative for activity monitoring and surveillance. Unlike cameras, radar is not typically subject to the same legal restrictions as it does not involve capturing visual images of individuals. This makes it a suitable option for use in public areas where privacy is a major concern, such as in places of worship or medical facilities. Another advantage of radar-based technology is that it is not affected by harsh weather conditions or darkness \cite{day_light_radar}, which can limit the effectiveness of camera-based surveillance systems. Radar is capable of detecting movements and activities even in low light or adverse weather conditions, making it a more reliable option for surveillance applications. In addition to its resilience in adverse weather conditions, radar also offers a higher range of detection compared to camera-based technology. This allows for a more comprehensive view of the monitored area, which is especially important in large public spaces.

Overall, while camera-based surveillance technology is widely used and has proven effective in many situations, it may not always be suitable or permissible in all scenarios. In such cases, radar technology can offer a viable alternative that can help address surveillance needs while respecting individuals’ privacy rights.

\begin{figure}[h] \flushleft 
\centering
\includegraphics[width=1\linewidth]{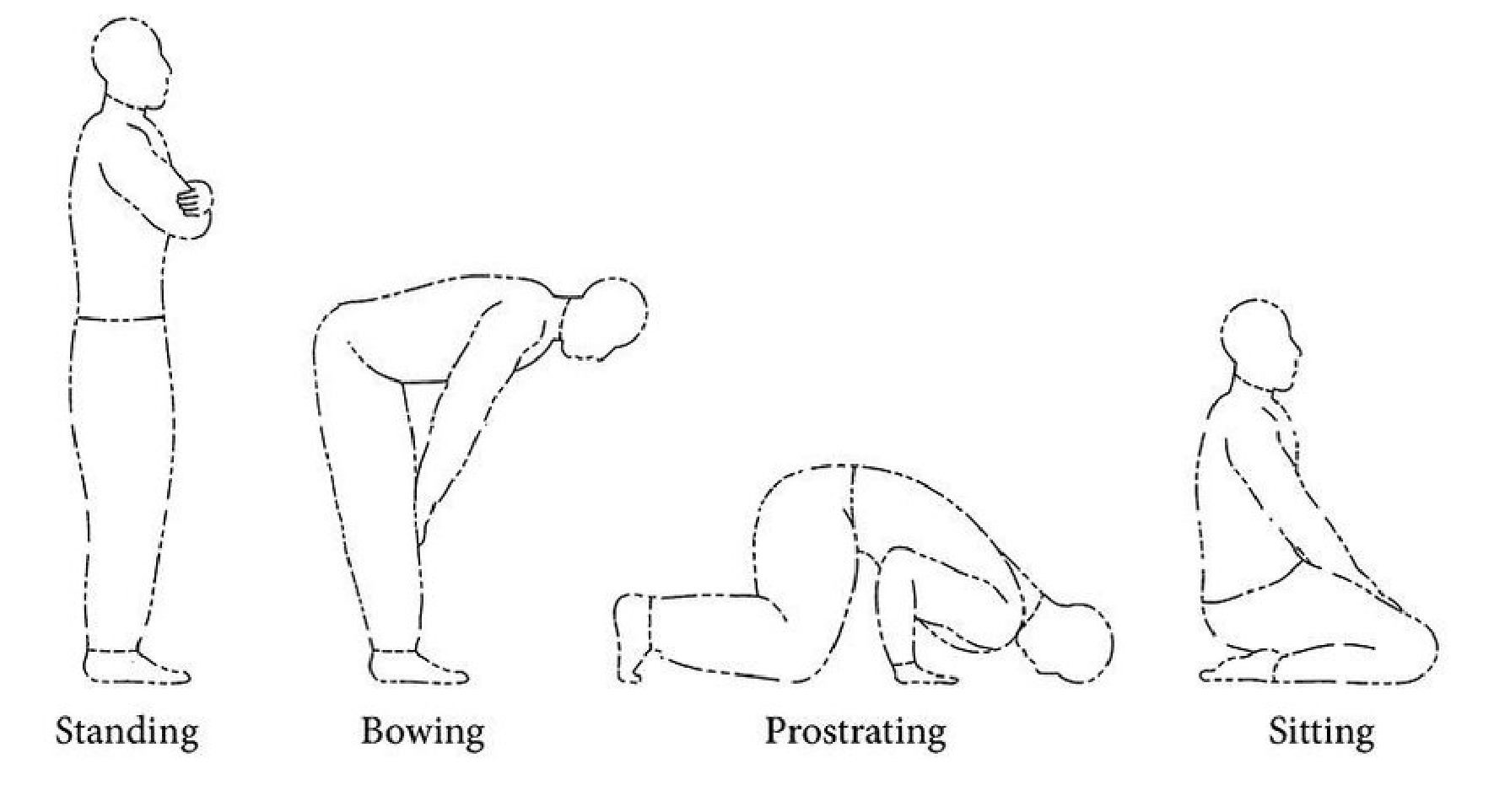}
\caption{The four basic activities in each rakat of a Salat are Standing, Bowing, Prostrating, and Sitting.} 
\label{fig:RakatMovements} 
\end{figure}

To demonstrate the effectiveness of radar-based activity monitoring, we have chosen to focus on an application for prayer tracking and assistance. The inspiration behind this project comes from the fact that salat (prayer) is one of the five pillars of Islam, with Muslims being obligated to perform it five times a day. Each salat involves repetitive cycles of four main activities, known as rakat, as illustrated in Fig. \ref{fig:RakatMovements}. Our proposed radar-based framework offers several benefits for prayer tracking and assistance systems. Unlike camera-based surveillance systems, our solution preserves individuals' privacy, since radar signals can penetrate walls and clothing and do not generate any images of individuals. Furthermore, our system can detect prayer activities with high accuracy, as well as identify any errors in the prayer movements, providing real-time feedback and assistance to users. 

During salat, it is possible for individuals to lose concentration and forget how many rakaat they have already performed. This is especially true for elderly individuals who may experience short-term memory loss, making it difficult for them to keep track of their prayer movements. Additionally, beginners often require a tutor to check their salat and ensure they are performing the movements correctly. Similarly, in congregational salat, latecomers may not know how many rakaat have already been performed by the Imam. To address these challenges, our radar-based framework provides a real-time device that can assist individuals during prayer. By accurately tracking the prayer movements, our system can provide users with real-time feedback on their performance, ensuring they perform the correct number of rakaat and movements. This can greatly benefit elderly individuals, beginners, and those participating in congregational salat. Furthermore, our system can provide real-time assistance to individuals who make mistakes during their prayer movements, allowing them to quickly correct their mistakes and continue with their salat without having to start over. This can be particularly useful for individuals who may have difficulty recalling the correct sequence of prayer movements. 

The related work in radar-based classification for human-related tasks is increasing, and there are some notable examples of its application in various areas. While there has been no prior research on radar-based prayer tracking and assistance, existing studies have focused on tasks such as posture classification \cite{postures_Mohamed, postures_friend, BodyCompassDina, forma_track}, vital sign monitoring \cite{Vital_signs_single_target,Vital_signs_multi_target, real_time_arm_radar}, and exercise monitoring \cite{Exercise_monitoring_single_person,Exercise_monitoring_multi_person, tangible_interactions}. Additionally, there have been applications of radar-based classification in the automotive industry, such as car, pedestrian, and cyclist classification, using both neural network and Convolutional Neural Network (CNN) based approaches \cite{CNN_radar_pedestrian,NN_radar_pedestrian,RudiSajid_Frontiers}. In most of these works, Frequency Modulated Continuous Wave (FMCW) radar is used. These studies demonstrate the potential of radar-based classification in various applications and highlight the need for further research in this area. 

The paper is structured as follows: Section II introduces the signal model. Section III describes the hardware used. Section IV provides an overall system description. Section V details the feature extraction and classification approaches. Section VI describes the dataset. Section VII discusses the classification results. Section VIII provides an investigation of the chosen classification architecture. Finally, Section IX concludes the work.

\section{Signal Model}
During prayer, a person performs a series of activities in a specific sequence. Each of these activities results in the movement of scatterers, which change their positions in terms of range and angles. The positions of these scatterers are unique to each activity, creating distinct patterns that can be analyzed.

Utilizing multiple antenna radar, we can estimate the range and angle of each scatterer if the distance between them exceeds the radar's resolution. Similarly, all scatterers whose angular distances surpass the radar's angular resolution can also be accurately estimated. This capability allows us to precisely track the positions of scatterers throughout the various prayer activities, enabling detailed and accurate activity detection. The radar's range and angular resolutions are given by
\begin{eqnarray}
\Delta R = \frac{c}{2B}, \notag \\
\Delta \theta = \frac{\lambda}{MNd} \notag \\
\end{eqnarray}
where $c$ is the velocity of light, $B$ is the bandwidth of the signal, $\lambda$ is the wavelength of the signal, $M$ is the number of transmit antennas, $N$ is the number of receive antennas, and $d$ is the distance between the adjacent antennas. Two scatterers can be uniquely identified if the distance between them is more then the range resolution or the angle between them is more than the angular resolution of the radar. Therefore, radar with multiple transmit and receive antennas can be used to localized different targets. 

During each prayer activity, the ranges and angular locations of the scatterers change in a distinct manner. Consequently, for each activity, the scatterers' ranges and angular locations combine to form a unique set. Each activity form a different set where some elements of the set overlap with other sets while most of them are different.
By combining radar capability to estimate the ranges and angular locations,  and passing unique sets corresponding to each activity as feature to a deep learning algorithm, we can develop a robust system for monitoring and classifying the activities performed during prayer. 

This approach can be extended to various other applications where individuals perform activities in a sequence. For instance, it can be utilized to monitor rehabilitation exercises, ensuring that patients perform their routines correctly and efficiently. By accurately tracking the movements and positions of scatterers, healthcare providers can gain valuable insights into a patient's progress and tailor their rehabilitation programs accordingly. 

Additionally, this technology holds significant potential in detecting specific activities or occurrences, such as falls or suspicious behavior in surveillance scenarios. By leveraging the precise range and angle estimations provided by multiple antenna radar, coupled with advanced deep learning algorithms, the system can identify and respond to unusual activities in real-time, enhancing safety and security in various environments.

{

To estimate the range, angle, and velocity, the FMCW radar mixes the transmitted and received signals coherently. The output of the mixer yields a single frequency signal corresponding to each target called a beat frequency signal given by 
\begin{equation}
    r(t) = \sum_{i=1}^P\alpha_i e^{j2\pi f_{b_i}t} \label{eq:beat}
\end{equation}
where $r(t)$ is the mixer output signal, $P$ is the total number of resolvable scatterer,  $\alpha_i$ is the complex gain while $f_{b_i} = \frac{BR_i}{cT}$ is the beat frequency of the scatterer $i$. As can be seen, the beat frequency depends on the range of scatterer $i$, denoted by $R_i$ and it can be easily identified by applying a Fast Fourier Transform (FFT) to the received signal. Each peak in the FFT represents the beat frequency corresponding to a scatterer at range $R_i$. Once beat frequency is estimated, the range of the scatterer $R_i$ can be calculated using \eqref{eq:beat}. 
Once the range of a scatterer is determined, its velocity can be estimated by applying FFT to the FFT peaks collected from each chirp corresponding to that scatterer within a frame. 
Finally, the angular position of the scatterer is determined by applying the FFT to the FFT peaks collected from each antenna associated with the scatterer \cite{stoicaRadarProcessing}. Unlike velocity estimation, where FFT samples are obtained from each chirp within a frame, for angular estimation, FFT samples are taken from the same chirp across all antennas.

In the following section, we will demonstrate the estimation of range, angular location, and velocity of different scatterers by utilizing a hardware module.


\section{Practical Demonstration}\label{Signal Model}
For the demonstration, we utilized Texas Instruments' off-the-shelf module AWR-1642 \cite{Radar_datasheet}. The top view of the module is shown in Fig. \ref{fig:awr1642_top}. This all-in-one module integrates an RF front-end, a digital signal processing (DSP) processor, and various peripherals for external connectivity. The module also has a set of different built-in diagnosis features to ensure robust operation. The RF front-end operates at a frequency of 77 GHz, with a maximum programmable bandwidth of 4 GHz for the transmitted signal. The module features two transmit and four receive antennas. The spacing between adjacent receive antennas is half the wavelength of the operating frequency, while the spacing between transmit antennas is twice the wavelength. This antenna configuration can provide an eight-element virtual steering vector to enhance spatial resolution \cite{stoicaRadarProcessing}. These multiple antennas enable us to determine the angular location of the target precisely. Each series patch antennas of the module emit/receive a narrow beam in the azimuth direction and a wide beam in the elevation direction. The linear vertical arrangement of these antennas enables the estimation of azimuth angles for various scatterers. This setup is particularly well-suited for tracking activities where scatterers move primarily in the elevation direction. For monitoring other activities, the orientation of the module can be adjusted accordingly.   
\begin{figure}[t] 
\begin{centering}
\includegraphics[width=0.6\textwidth]{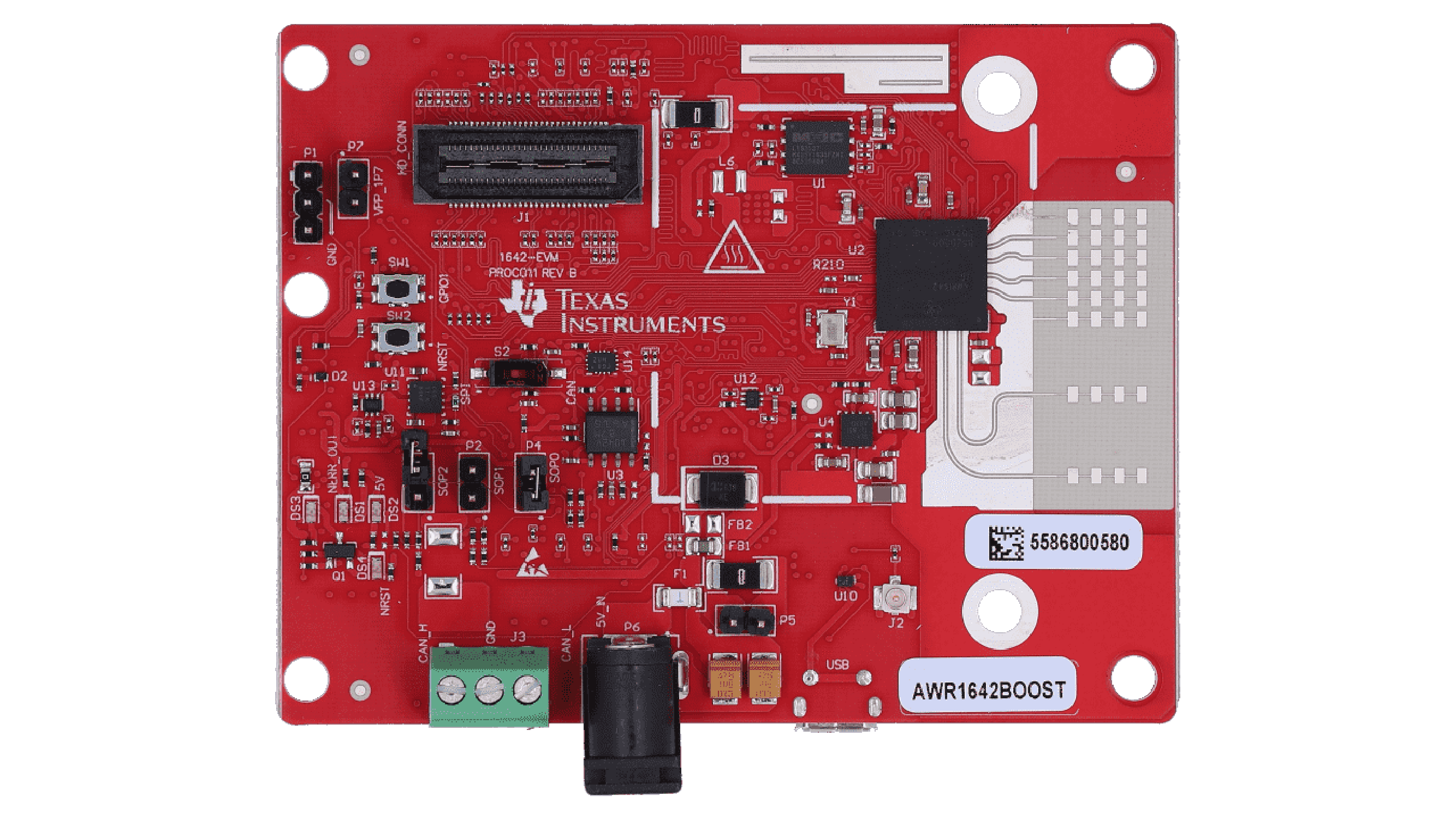}
     \caption{The actual orientation of the TI AWR-1642 module (Front View) to capture the data using 2-Tx and 4-Rx antennas. The module in this position estimates the range and elevation angles of different scatterers.}
    \label{fig:awr1642_top}
\end{centering}
\end{figure}
\\
To initiate the tracking, an FMCW signal, also known as a chirp signal, of a predefined duration is transmitted multiple times in a frame. The frame duration defines the velocity resolution. The extreme estimation capabilities of the module, corresponding to some specific programmable parameter values, are outlined in Table \ref{tab:awr_table}. These key parameter values are highlighted in {\bf bold letters} for clarity and emphasis.
\begin{table}[ht] 
\caption{AWR-1642 Texas Instrument radar module parameter configuration.}
\begin{center}
\begin{tabular}{|c|c|}
    \hline
    Parameter & Value\\
    \hline
    ${\bf B}$ & 4$\;$GHz \\
    \hline
    {\bf Frame duration} & $50$ ms\\
    \hline
    Azimuth Resolution & $15 ^\circ$ \\
    \hline
    Maximum unambiguous Range & $2.41$ m \\
    \hline
    Range Resolution & $3.75$ cm \\
    \hline
    Max. Velocity  & 1$\;$m/s\\
    \hline
    Velocity Resolution  & 0.13$\;$m/s\\
    \hline
    \end{tabular}
    \end{center}\label{tab:awr_table}
\end{table}
Additionally, the board offers multiple interfaces for data collection, such as USB for point cloud data and a low-voltage differential signaling bus for transferring raw data, making it convenient for researchers. The board is priced at approximately \$300, making it an affordable option for research purposes.
The use of an off-the-shelf module not only simplifies the hardware setup but also allows for quick prototyping and testing of the proposed system. The TI AWR-1642 module provides the necessary capabilities to accurately estimate the range, angle, and velocity of the target. Its small form factor and low power consumption make it an ideal choice for our proposed system.
\section{FRAMEWORK}
The radar-based classification requires numerous degrees of freedom. Initially, we configure and position the radar to ensure robust operation. Next, received raw data is cleaned, different features are extracted from it, and the raw data captured from an activity is classified. The overall short description of the algorithm is illustrated in the block diagram shown in Fig. \ref{img:framework_diagram}.



\begin{figure}[h]
\begin{tikzpicture}[node distance=3cm, auto] 
\node [draw, rectangle, minimum width=2cm, minimum height=1cm,  fill=pink, line width=1.25pt] (block1) {\parbox{3.6cm}{~~Parameter Initialization and Module Positioning}};
\node [draw, rectangle, below of=block1, yshift=0.5cm, minimum height=1cm,fill=green!20, line width=1.25pt] (block3) {\parbox{3.3cm}{~~Posture Tracking and ~~Feedback Generation}};
\node [draw, rectangle, right of=block1, xshift=2cm, minimum height=1cm, fill=green!20, line width=1.5pt] (block2) {\parbox{3.3cm}{~~Raw Data Processing and Feature Extraction}};
\node [draw, rectangle, below of=block2, yshift=0.5cm, minimum height=1cm,fill=pink, line width=1.25pt] (block4) {\parbox{3.6cm}{~~Features to Imaging and Activity Classification}};

\draw[->, line width=1.5pt] (block1) -- (block2);
\draw[->, line width=1.5pt] (block2) -- (block4);
\draw[->, line width=1.5pt] (block4) -- (block3);
\end{tikzpicture} 
\caption{Framework for system development.}
\label{img:framework_diagram}
\end{figure}
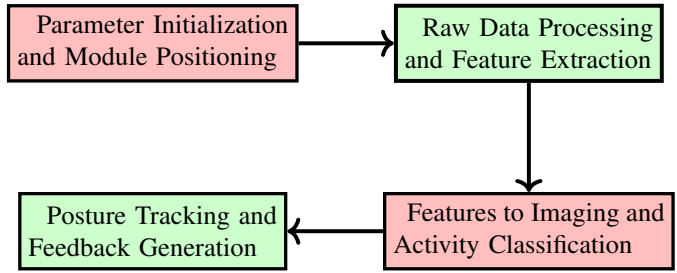



For radar positioning, we selected the most convenient option for the user perspective and data collection point of view. The sensor is placed on the ground in front of the person, approximately 1.5 meters away, with an elevation angle of about 60 degrees as shown in Fig. \ref{fig:Posture_schematic}. Both the distance from the radar module to the person and the angular tilt of the module can vary within certain limits.
This positioning might distort the expected range-angle scatterers image of a posture due to the varying height of the person and the lengths of different body parts. However, we anticipate that the reflected signal includes multiple scatterer points, collectively forming a unique image for each posture. Therefore, slight variations in the module's positioning and the person's height do not affect the classification.

Note that our algorithm is designed to avoid classification during the transition between postures. This is achieved by avoiding classification whenever there is a Doppler shift in the raw data.

\begin{figure}[htbp]
\centerline{\includegraphics[width=0.2\textwidth]{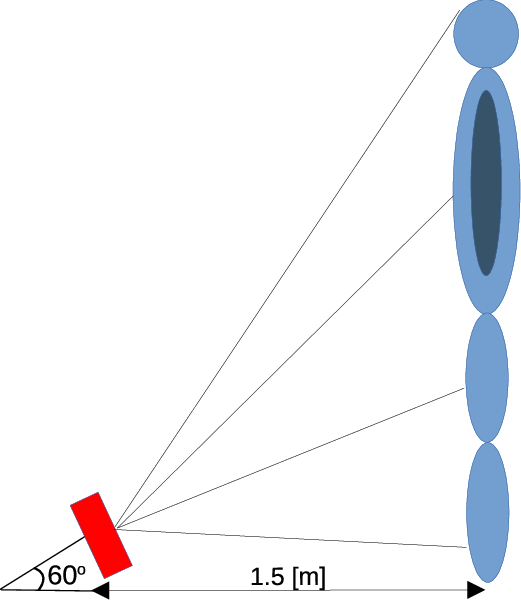}}
\caption{Radar setup example in front of the person. Radar is denoted in red.}
\label{fig:Posture_schematic}
\end{figure}

Conventional radar processing chain on fast- and slow-time samples yields an array of detected scatters with unique range and angle amplitude peaks along with Doppler shift values depending on the posture. Examples of detected scatterer points for four different postures of prayer are provided in Fig. \ref{fig:ExamplePointCloud}. There are many static and random noise scatters present in this data, so we need to filter them out. 

\begin{figure}[ht]
    \begin{minipage}[b]{0.5\linewidth}
        \centering
        \includegraphics[width=\linewidth]{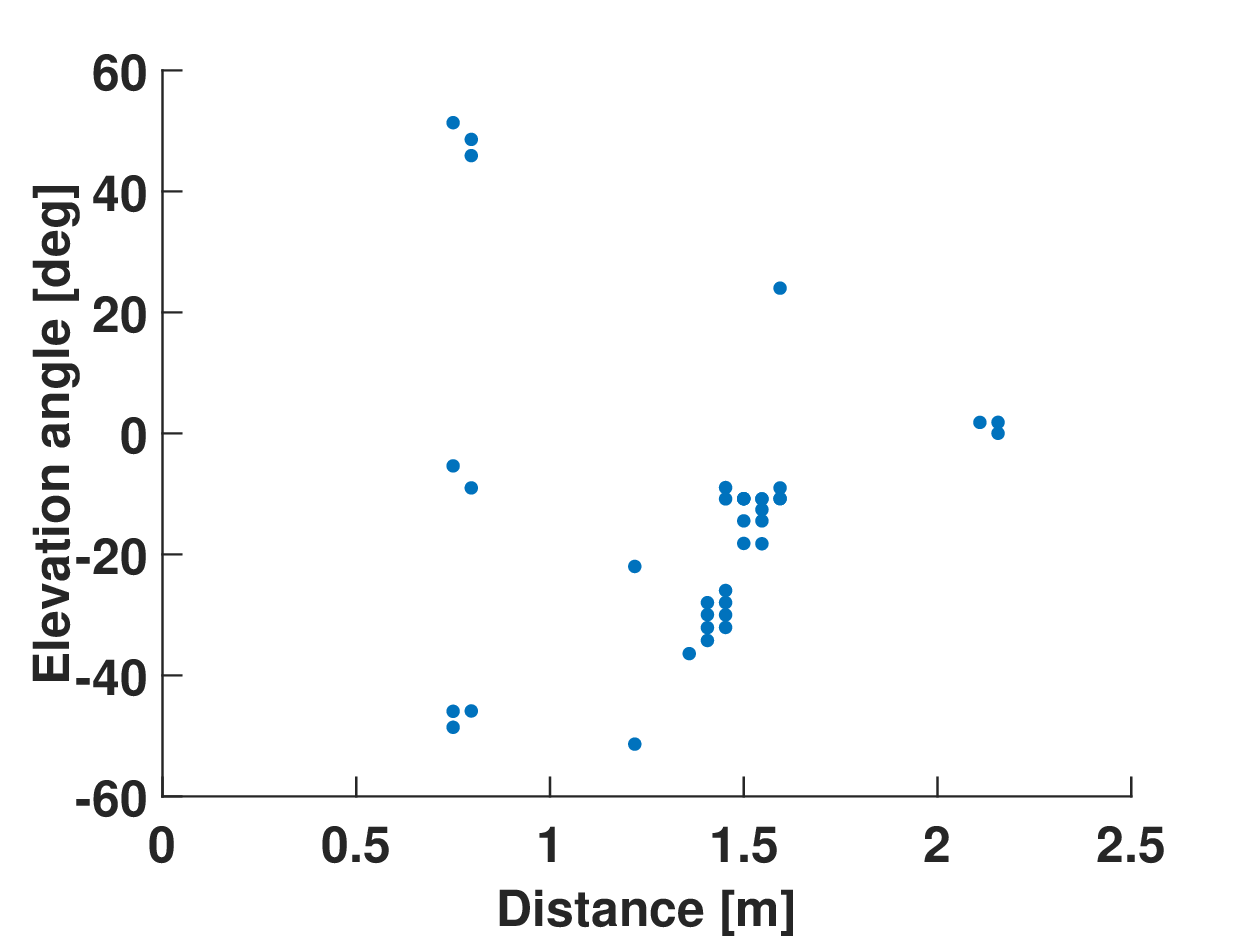}
        \caption*{(a)} 
        \vspace{4ex}
    \end{minipage}
    \begin{minipage}[b]{0.5\linewidth}
        \centering
        \includegraphics[width=\linewidth]{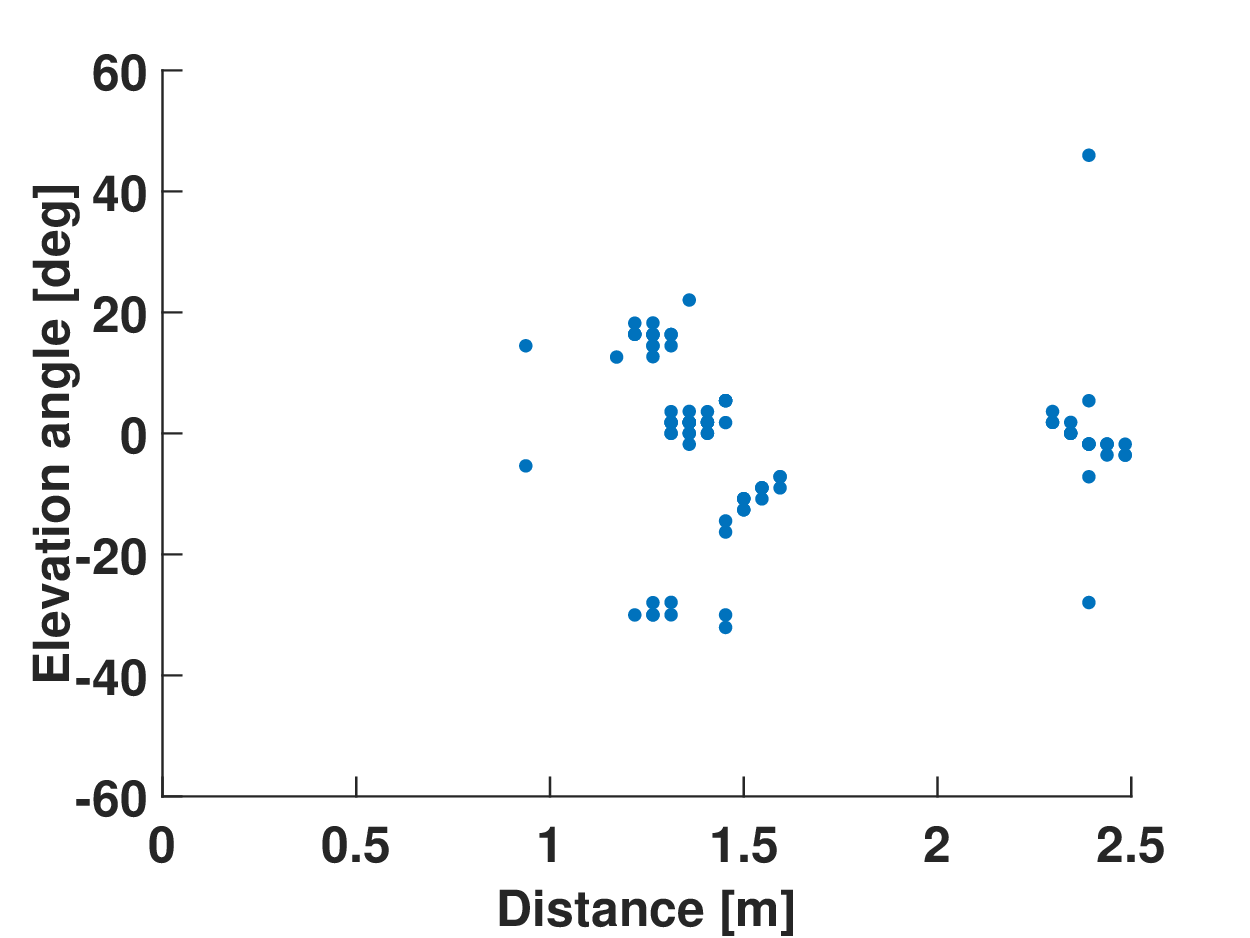}
        \caption*{(b)}
        \vspace{4ex}
    \end{minipage}

    \begin{minipage}[b]{0.5\linewidth}
        \centering
        \includegraphics[width=\linewidth]{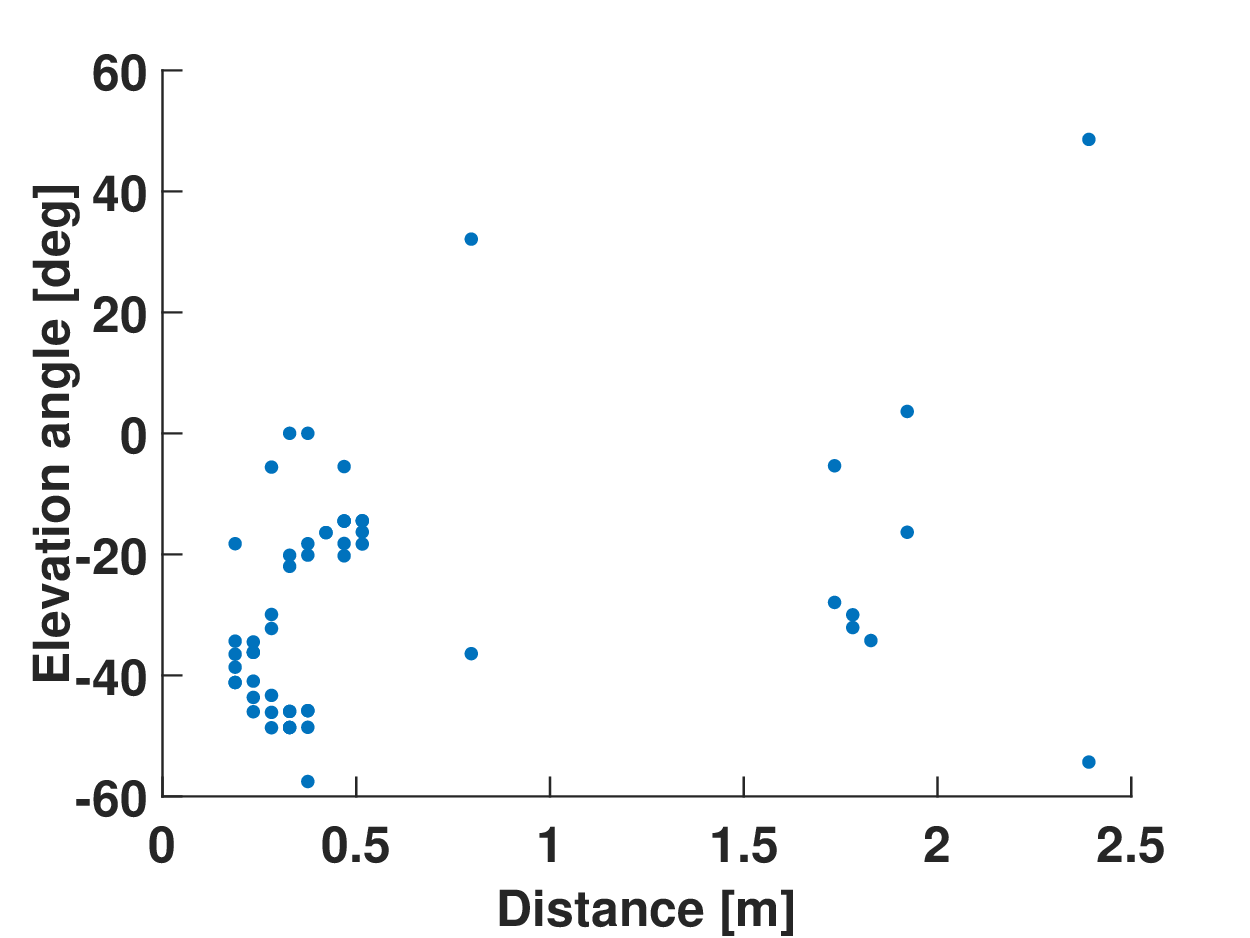}
        \caption*{(c)}
        \vspace{4ex}
    \end{minipage}
    \begin{minipage}[b]{0.5\linewidth}
        \centering
        \includegraphics[width=\linewidth]{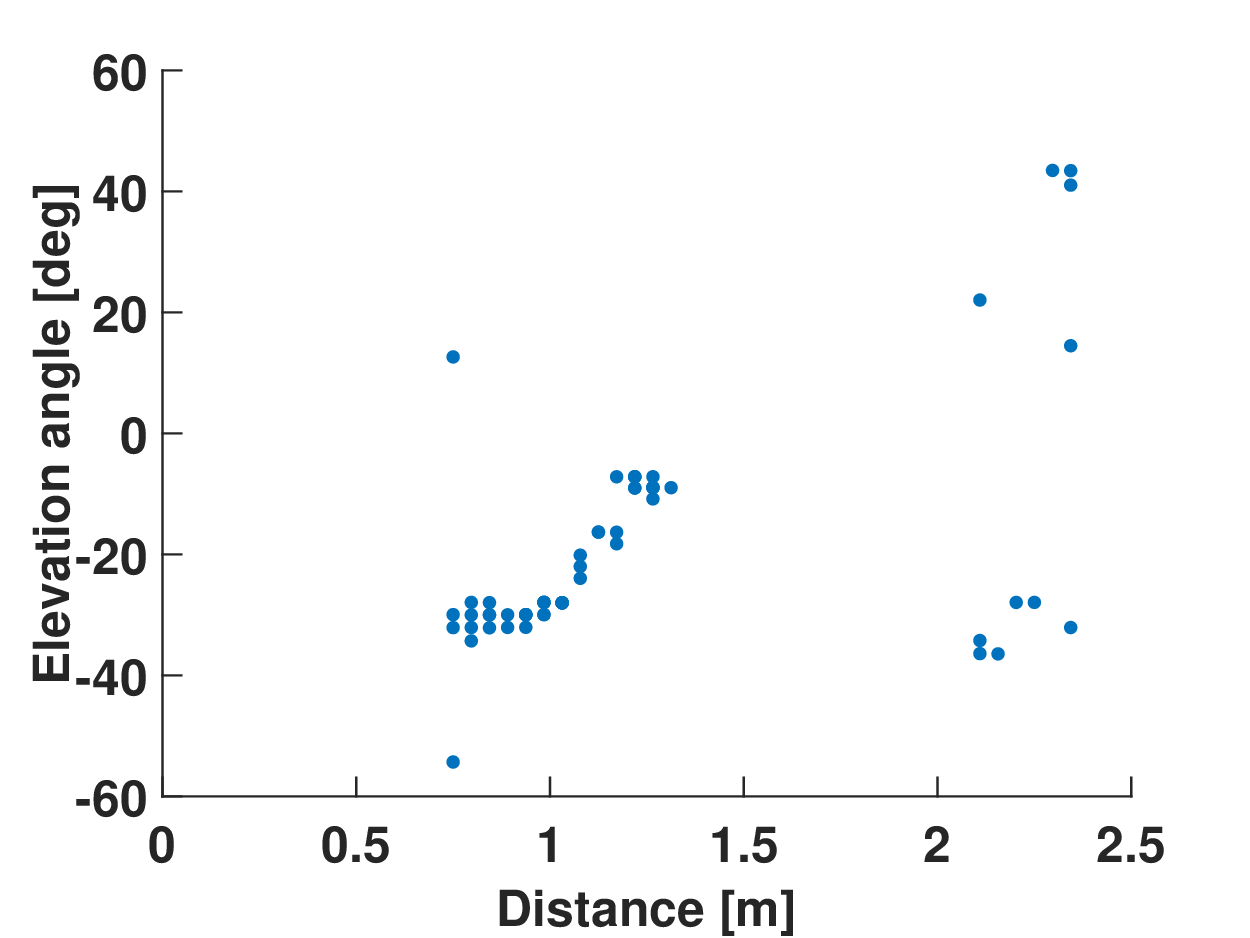}
        \caption*{(d)}
        \vspace{4ex}
    \end{minipage}
    \caption{Example of scatters for (a) standing, (b) bowing, (c) prostrating, and (d) sitting position after radar processing.}
    \label{fig:ExamplePointCloud}
\end{figure}

Data preprocessing begins with transforming scatter plot into a two-channel image representation. Each pixel in the resulting 35 by 40 pixel image corresponds to a specific spatial element defined by range and angle. One channel of the image captures peak reflected amplitude values, while the other channel captures Doppler values.

Subsequently, a scaling and centering process, along with low-pass filtering techniques, is applied to multiple scatters  to reduce random noise and highlight points of interest. An illustrative output image for standing and bowing is depicted in Fig. \ref{fig:kyam_image} and \ref{fig:ruku_image}.

\begin{figure}[ht]
    \begin{minipage}[b]{0.5\linewidth}
        \centering
        \includegraphics[width=\linewidth]{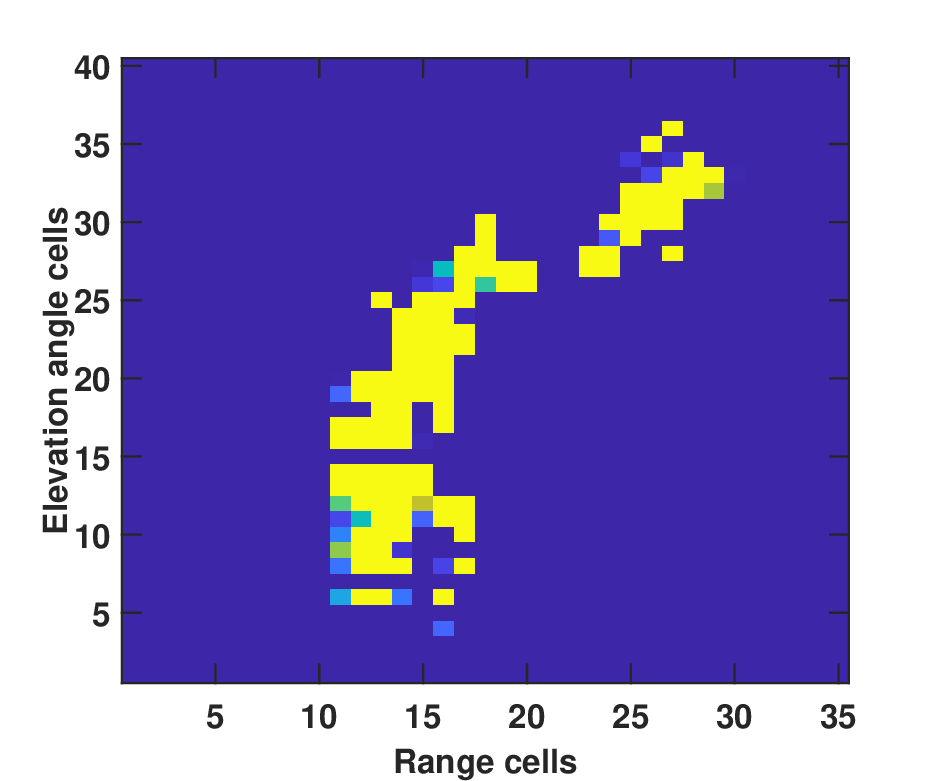}
        \caption*{(a)}
        \vspace{4ex}
    \end{minipage}
    \begin{minipage}[b]{0.5\linewidth}
        \centering
        \includegraphics[width=\linewidth]{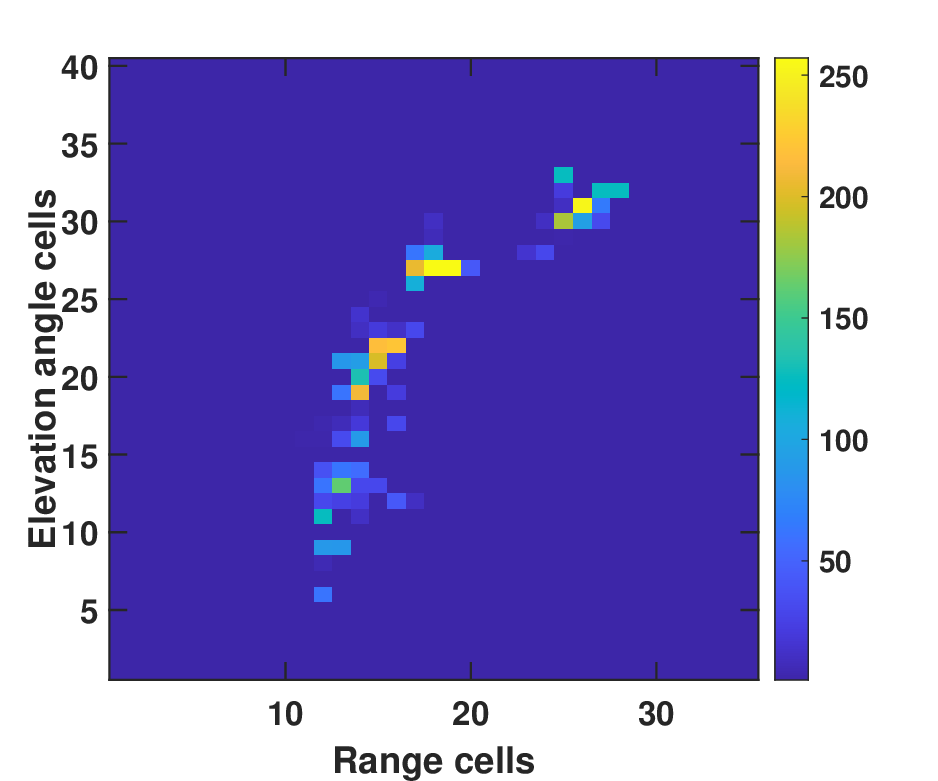}
        \caption*{(b)}
        \vspace{4ex}
    \end{minipage}
    \caption{\textit{Standing} position transformed image of (a) reflection coefficient and (b) Doppler shift channels.}
    \label{fig:kyam_image}
\end{figure}
\begin{figure}[ht]
    \begin{minipage}[b]{0.5\linewidth}
        \centering
        \includegraphics[width=\linewidth]{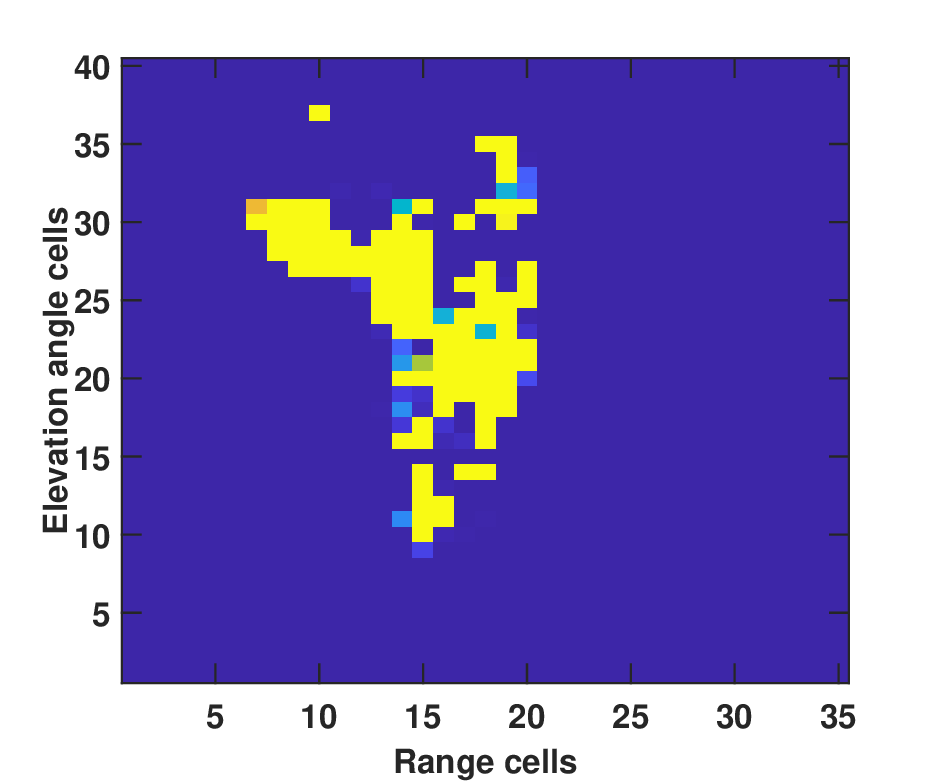}
        \caption*{(a)}
        \vspace{4ex}
    \end{minipage}
    \begin{minipage}[b]{0.5\linewidth}
        \centering
        \includegraphics[width=\linewidth]{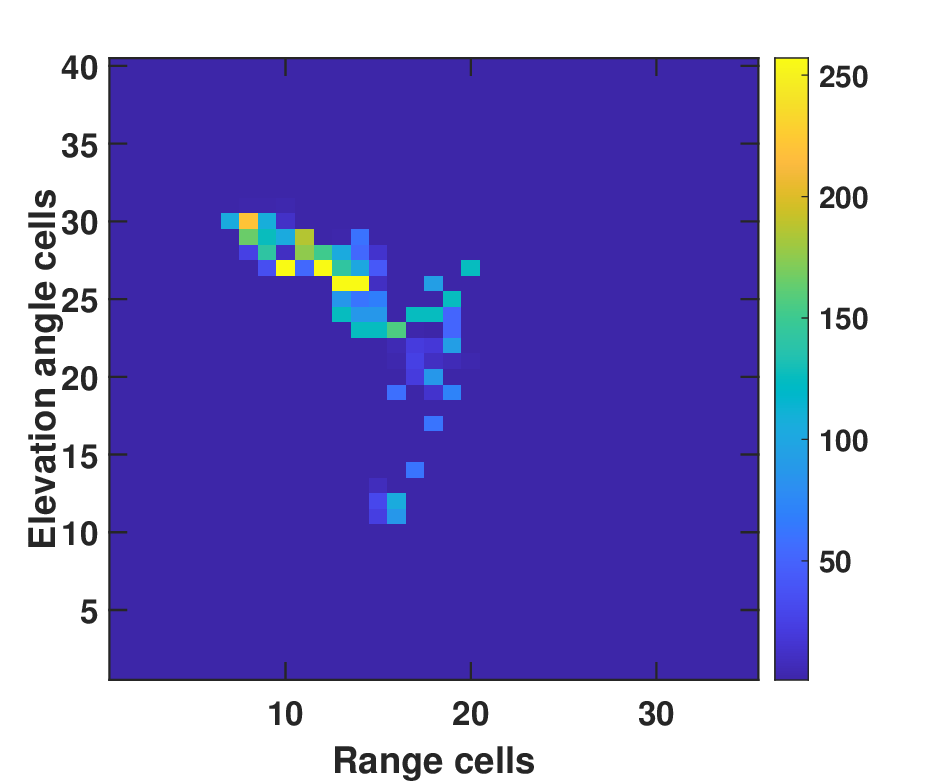}
        \caption*{(b)}
        \vspace{4ex}
    \end{minipage}
    \caption{\textit{Bowing} position image of (a) reflection coefficient and (b) Doppler shift channels.}
    \label{fig:ruku_image}
\end{figure}

After preprocessing, we can clearly distinguish between different postures in the images. For example, in the standing posture, the lower part of the reflection-coefficient channel image represents the lower part of the person, while the upper part of the image represents the upper part of the person. This aligns with our expectations: the lower part of the person is closer to the radar module and has lower angle values, whereas the upper part is relatively farther away from the radar module and has higher angles. Both of these observations can be interpreted by examining the image. On the other hand, the Doppler shift channel image shows negligible values for most of the scatterers. Only a few scatterers in the middle display some Doppler shift, which can be attributed to chest movements caused by heartbeat and breathing. 

Conversely, in the bowing posture, as expected, we observe the head positioned closer to the radar compared to the body. Similar to standing posture, bowing posture can be easily interpreted in Fig. \ref{fig:ruku_image}. Although not shown here, other postures can also be easily interpreted from their images, each exhibiting unique characteristics.


The next step involves extracting features from the processed data and applying a classification algorithm. The following sections will provide a detailed description of these topics.

To provide feedback for users, a control logic was designed using a Moore finite state machine. Each state represents a different prayer position. Transitions between states occur upon detecting changes in the observed posture. To enhance robustness and keep the acceptable response time, a buffer stores the last 3 predictions. The ultimate decision regarding the change in prayer position is based on the dominant prediction in the buffer. 
The correct sequence of postures in a prayer that a person must follow is presented in Fig. \ref{fig:prayer_seq}.  
\begin{figure}[htbp]
\centerline{\includegraphics[width=0.5\textwidth]{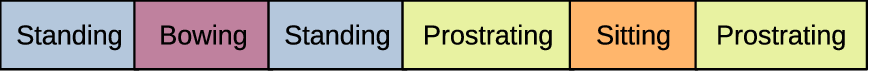}}
\caption{Expected prayer sequence.}
\label{fig:prayer_seq}
\end{figure}

If we assume the user follows the prayer sequence correctly, we can improve classification accuracy by incorporating prior knowledge about probabilities of transitions. For instance, knowing that a person is in a standing position, there is a high probability they will move to bowing, a very small probability they will move to prostration, and zero probability they will move to sitting. However, this approach has a drawback: the system may not detect errors in the subsequent prayer sequence.

After considering the factors discussed above, we have determined that a desired classification accuracy of above 90 percent is necessary for the classification algorithm.

\section{Feature extraction and classification}
To offer a comprehensive comparison, we evaluated various feature extraction and classification algorithms. 
Feature extraction approaches are provided below:
\begin{itemize}
    \item \textit{Common Sense Feature Extraction:} This approach is based on intuitively selecting relevant data from observed dataset. We identified 20 key features from the processed point cloud, including maximum angle, mean range, and the range associated with the maximum angle.
    \item \textit{Principal Component Analysis (PCA):} In this approach we applied singular value decomposition on training data, and chose the 300 most important image representatives, also known in image recognition as eigenfaces. Given unknown, flatten to vector image $\textbf{I}$, we can decompose it using basis of eigenfaces stored in matrix $\textbf{U}$ with some weight vector $\textbf{w}$ given by \cite{PCA}
    \begin{equation} \label{eq:efaces}
        \textbf{w} = (\textbf{U}^T\textbf{U})^{-1}\textbf{U}^T\textbf{I}.
    \end{equation}
    The vector $\textbf{w}$ resulting from this decomposition serves as the feature vector for further classification.
    \item \textit{Sparse Image Representation:} This approach shares similarities with PCA, yet differs in several important aspects. Based on training image pool we prepared dictionary $\textbf{D}$, which is collection of basis images to reproduce any other image. Unlike PCA, our objective is to maximize the size of this dictionary to achieve a sparse decomposition. For any unknown flattened image $\textbf{I}$, we decompose it by solving the following optimization problem \cite{Sparse_signal_processing}
     \begin{equation} \label{eq:l0}
    \begin{aligned}
     & \min_{\textbf{x}}\quad \lVert \bf{x} \rVert_{0} \quad \\
    &\textrm{s.t.} \quad \textbf{D}\bf{x} = \bf{I} ,
    \end{aligned}
    \end{equation}
    where $\textbf{x}$ is a binary decomposition vector and $\lVert \bf{x} \rVert_{0}$ is so-called $l_0$-norm.
    
    To obtain approximate solution of optimization problem \eqref{eq:l0}, we employed orthogonal matching pursuit algorithm. Interestingly, images corresponding to similar positions often share common dictionary elements during decomposition. This allows us to distinguish between different positions by knowing which dictionary elements represent a specific image $\textbf{I}$. In contrast to PCA, decomposition vector $\textbf{x}$ is longer but binary. This characteristic simplifies the classification algorithm but at higher cost of image decomposition.
\end{itemize}

Once, features are extracted, we used the following approaches to classify different postures of the prayer \cite{Furkranz2010}.
\begin{itemize}
    \item \textit{Decision Trees:} This algorithm aligns with human intuition for feature-based classification and is well-suited for common-sense feature extraction.
    \item \textit{Multiclass Logistic Regression:} This classification approach is effective for input vector of features and is particularly well-suited for use with PCA feature extraction.
    \item \textit{Minimum Euclidean Distance:} This method, which classifies data based on the closest class representatives, is computationally cheaper than other approaches for PCA features.
    \item \textit{Dense Neural Network:} Although this method has the highest complexity, it offers the potential for additional feature extraction followed by classification.
    \item \textit{Vector Support Comparison:} This classification method involves comparing the support of input images with representative images from various classes. The class to which an image belongs is determined by how closely its support align with those of the representative images, which is measured by the number of similar basis images.
    \item \textit{CNN:} This approach has become standard for image classification in recent years. Its capability to learn and extract features from data aligns perfectly with considered problem. From the variety of available architectures, we decided to study two well-known ones: ResNet \cite{resnet_paper} and SqueezeNet \cite{squeezenet_paper}. By leveraging both, we aim to investigate the balance between accuracy (ResNet) and computational efficiency (SqueezeNet).
\end{itemize}

Also, in order to reduce complexity we can apply PCA on input data and apply smaller images to CNN. This process is called compression. In our research, we resized the original images from 35 by 40 pixels to 20 by 21 pixels to study how compression affects different CNN setups.

\section{Dataset description}
For the considered problem, we utilized a dataset of approximately 40,000 radar shots for the four postures in prayer. This dataset was collected across 10 scenarios in various environments. Each individual exhibits unique dimensions and movement patterns, with heights ranging from approximately 170 cm to 185 cm. Different instances of the same individual vary in radar positioning, environmental conditions, and prayer posture.

There are three distinct scenarios to evaluate the classification accuracy. Each scenario provides valuable insights into the algorithm's performance:
\begin{enumerate}
    \item Randomly split the whole dataset into training and test sets to evaluate the accuracy of classification for all known information such as individual parameters, prayer style, and environmental conditions.
    \item Use nine scenarios for training and reserve one for testing. Ensure that the testing scenario features the same individual as seen in the training data. This checks the algorithm's ability to classify a familiar person under different circumstances.
    \item Select nine scenarios for training and keep one entirely separate for testing, ensuring it does not appear in the training data. This evaluates the algorithm's capability to classify completely unfamiliar individuals.
\end{enumerate}

From a real-world implementation perspective, the third approach holds particular importance, although all three are essential for understanding the behavior of classification algorithms.

\section{Classification results}

Classification accuracies for each data division scenario across different approaches are provided in the Fig. \ref{fig:dds1}-\ref{fig:dds3}.

\begin{figure}[htbp]
\centerline{\includegraphics[width=0.5\textwidth]{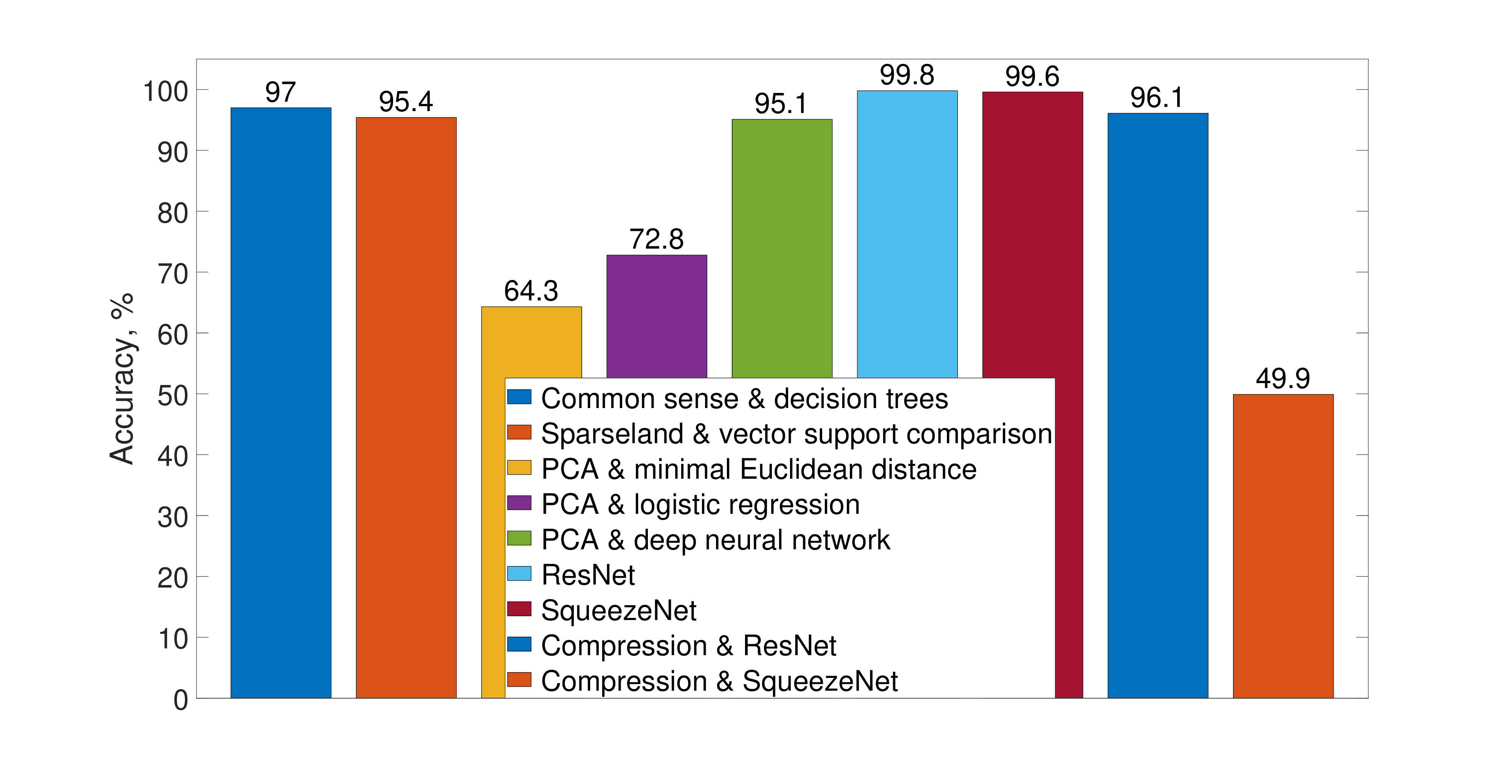}}
\caption{Test accuracy for the first data division scenario.}
\label{fig:dds1}
\end{figure}
\begin{figure}[htbp]
\centerline{\includegraphics[width=0.5\textwidth]{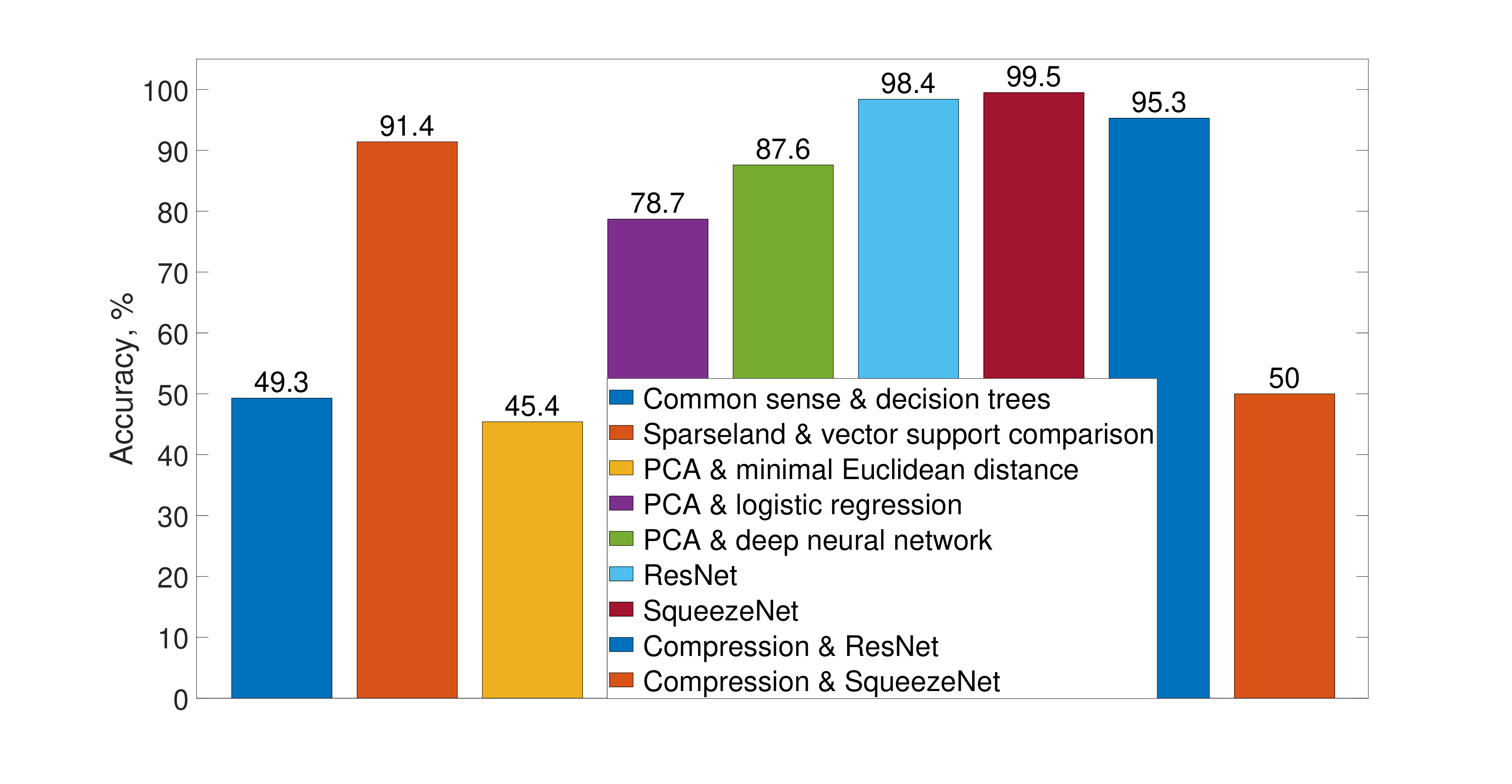}}
\caption{Test accuracy for the second data division scenario.}
\label{fig:dds2}
\end{figure}
\begin{figure}[htbp]
\centerline{\includegraphics[width=0.5\textwidth]{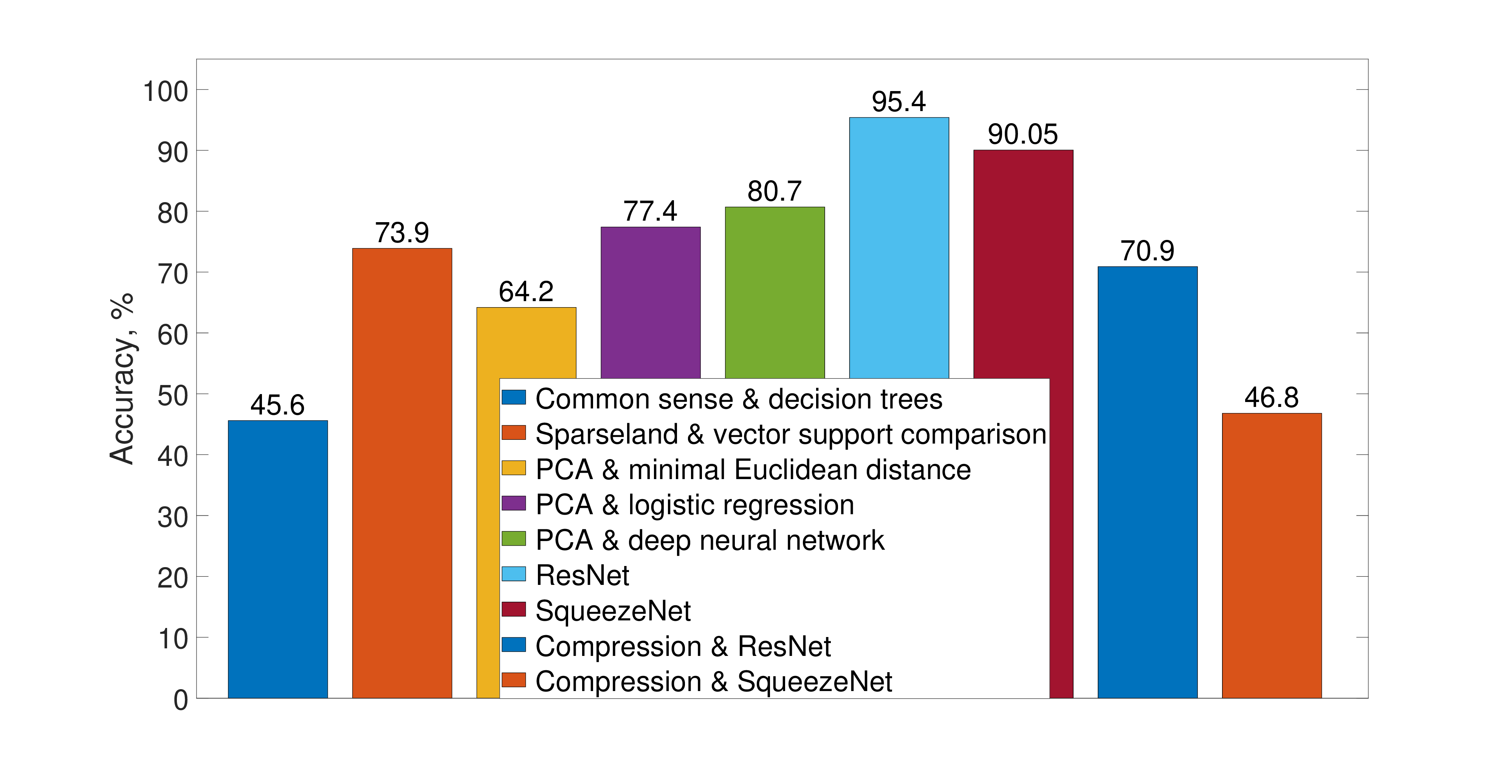}}
\caption{Test accuracy for the third data division scenario.}
\label{fig:dds3}
\end{figure}

In the context of naive feature extraction using decision tree classification, we observe that decision trees yield high accuracy in the first scenario. However, in the second scenario, there is a notable decrease in accuracy, indicating limited generalization ability and susceptibility to minor data perturbations. This suggests that the approach may not generalize well across different datasets. Furthermore, in the third scenario, there is negligible distinction between known and unknown cases, underscoring the model's inability to effectively differentiate between these scenarios.

When employing sparseland feature extraction followed by vector support comparison, we find consistent high accuracy in both the first and second scenarios, demonstrating robust generalization to known types of persons and resilience to data perturbations. However, in the third scenario, while there is some level of generalization observed, the accuracy achieved is insufficient for the given task.

PCA-based feature extraction demonstrates varied effectiveness depending on the classifier structure chosen. The simplest classifier relying on Euclidean distance yields poor results, indicating its limited discriminatory power. In contrast, more complex classifiers exhibit improved classification quality. Notably, neural networks achieve high accuracy, even when presented with unknown data.

An intriguing observation arises from the second data scenario where we observe a decline in testing performance when using the minimum Euclidean distance classifier. This suggests that the specific information in the unknown dataset may not be adequately captured by PCA. One potential solution could involve increasing the amount of training data. Alternatively, employing CNN classifiers, known for their robustness based on current results, may offer a more effective solution. To reduce training complexity, we employed transfer learning by leveraging a pre-trained networks from the ImageNet dataset.

When employing CNN architectures without image compression, both ResNet and SqueezeNet consistently yield excellent results. However, upon applying image compression, we observe a decline in performance: ResNet shows reduced accuracy in the third data scenario, while SqueezeNet experiences a significant drop in accuracy. This decrease is exacerbated by SqueezeNet's inherent dimensionality reduction within its architecture, which, coupled with PCA compression, leads to substantial information loss. 

Considering the balance between complexity and accuracy, we recommend utilizing the SqueezeNet architecture without image compression.

\section{Architecture investigation}
In this section, we investigate SqueezeNet architecture. A depiction of this network tailored for a 2-channel input image and incorporating 3 layers of fire modules, is illustrated in Fig. \ref{fig:SqueezeNet_arch}.

\begin{figure}[htbp]
\centerline{\includegraphics[width=0.5\textwidth]{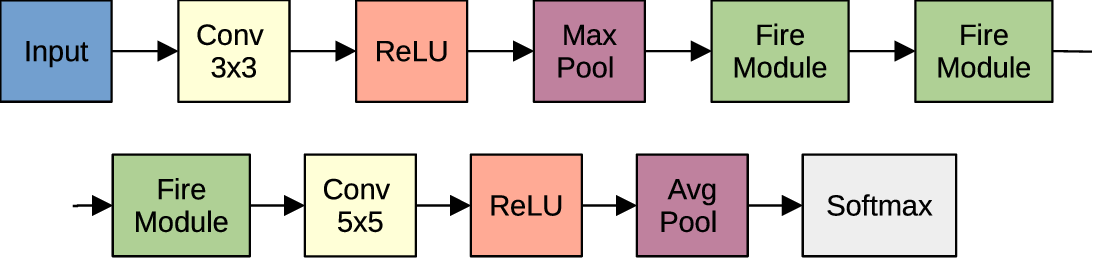}}
\caption{SqueezeNet structure.}
\label{fig:SqueezeNet_arch}
\end{figure} 

Structure of each fire module is presented in Fig. \ref{fig:fire_module}.

\begin{figure}[htbp]
\centerline{\includegraphics[width=0.5\textwidth]{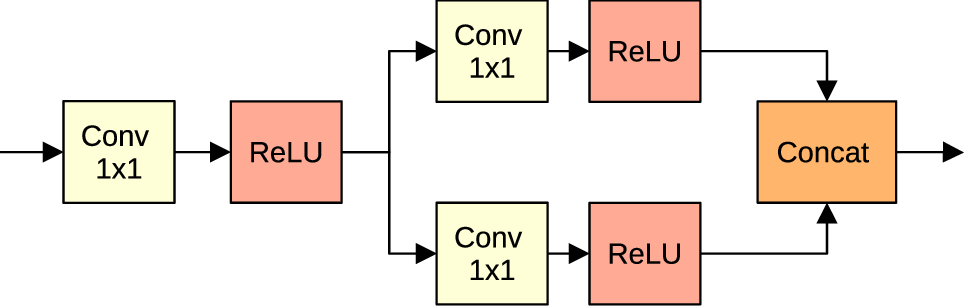}}
\caption{Fire module for SqueezeNet.}
\label{fig:fire_module}
\end{figure} 




Since we are using a 2-channel image, we investigated the performance when using only one input channel (either reflection amplitude or Doppler). Our findings indicate that the reflection channel contains sufficient information to achieve the same accuracy as using both channels, whereas the Doppler channel provides less informative results, as observed from the comparison below.

To train the single-channel networks, we initialized with a pre-trained model from the two-channel setup and adapted the first convolutional layer accordingly. Our objective was to attain results at least comparable to those of the two-channel model.

For comparative analysis, we measured accuracy relative to the two-channel model, setting its accuracy as 1. The performance comparison across these scenarios is depicted in Fig. \ref{fig:SqNet_comparison}.

\begin{figure}[htbp]
\centerline{\includegraphics[width=0.5\textwidth]{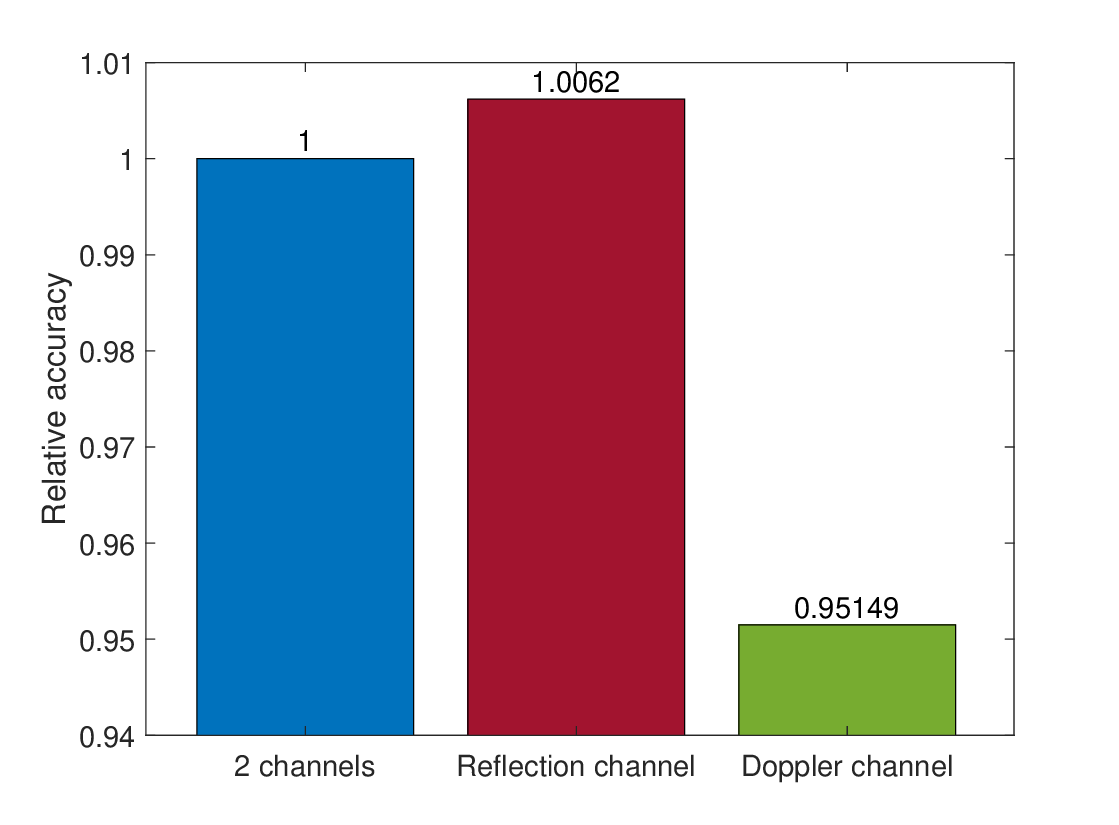}}
\caption{Comparison of SqueezeNet accuracy based on varying numbers of input channels, relative to performance with 2 channels.}
\label{fig:SqNet_comparison}
\end{figure}

We have observed that the reflection channel alone contains sufficient information to achieve the desired classification accuracy. While utilizing the reflection channel might marginally enhance performance compared to using both channels, any difference is minimal and could be attributed to random variability during training. When deciding between channels, computational complexity favors the use of just the reflection channel due to its lower resource requirements.

For a real-world demonstration, we conducted an additional experiment involving a new individual in a different environment. This person was asked to perform prayers with a prayer assistant, successfully completing the task multiple times with accurate responses from the fully operating system.
\\\\
{\bf Demonstration Video:}\\ 
We have prepared a demonstration of this work. In the real-time demonstration, the developed system displays the current posture of the person using doodle pictures, with the posture name shown in black text at the top of the doodle. The next posture is indicated in red text. After each activity, the system updates to display the next posture. The person performing the prayer can view the screen for guidance, ensuring they follow the correct sequence of postures.
The demonstration video can be viewed at the following link: \url{https://youtu.be/PnpGQZWqCr4}.

\section{Conclusion}
This paper presents an approach to classify static human postures derived from detected point clouds using a conventional FMCW radar processing scheme. 
To illustrate the framework described, we apply it to the prayer tracking application. We introduce a framework designed to address problem, with brief descriptions of each step. In order to handle with noise, we propose to use filtering and additional preprocessing to obtain images corresponding to current position. We compared the performance of different machine learning based methods including deep convolutional neural networks. Our findings indicate that transforming data into images using only the reflection coefficient channel, coupled with the SqueezeNet classification algorithm, achieves acceptable performance and ensures reliable system operation.

As a future direction, we can explore the potential of the system for dynamic recognition, similar to video recognition systems. In order to move in this direction, it is only required to replace the classification algorithm with a more complicated one, and ensure employing a Doppler channel, since for dynamic recognition, Doppler information plays a more important role than for static recognition. Other processing steps could be the same.

\bibliographystyle{ieeetr}
\bibliography{References}

\end{document}